\documentclass[10pt,twocolumn,twoside] {IEEEtran}
\usepackage{setspace}
\newcommand{\Z}{\mathcal{Z}}
\usepackage{ifthen}
\usepackage {ftnxtra}
\usepackage{placeins}
\usepackage{algpseudocode}

\usepackage{amsmath,graphicx}
\usepackage[english]{babel}
\usepackage[latin1]{inputenc}
\usepackage{tikz}

\newcommand{\displaycomments}

\ifdefined\displaycomments
\newcommand{\note}[1]{\textcolor{red}{\emph{#1}}}
\else
\newcommand{\note}[1]{}
\fi

\usepackage{xspace}
\usepackage{amsmath}
\usepackage{amsthm}
\usepackage{amssymb}
\usepackage{graphicx}      
\usepackage{algorithmicx} 
\usepackage{cite} 

\usetikzlibrary{arrows}
\usepackage{pgfplots}

\DeclareMathOperator*{\argmin}{argmin}

\newcommand{\E}{\mathop{\mathbb E}}

\renewcommand{\t}[1]{\mathrm{T}#1}
\newcommand{\N}{\mathcal{N}}

\newcommand{\diag}{\operatorname{Diag}}
 
\usepackage{cite} 
\usepackage{stfloats} 
\usepackage{amsmath}

\renewcommand{\d}[1]{\;\mathrm{d}#1}
\renewcommand{\t}[1]{\mathrm{T}#1}

\newcommand{\tr}{\operatorname{Tr}}

\newcommand{\IW}{\mathcal{IW}}

\newcommand{\W}{\mathcal{W}}

\renewcommand{\d}[1]{\;\mathrm{d}#1}
\renewcommand{\t}[1]{\mathrm{T}#1}

\usepackage{subfigure}
\usepackage{booktabs}
\usepackage{algpseudocode}

\title{Approximate Bayesian Smoothing with Unknown Process and Measurement Noise Covariances}
\author{Tohid Ardeshiri, Emre \"{O}zkan, Umut Orguner, Fredrik Gustafsson
\thanks{T. Ardeshiri, E. \"{O}zkan and F. Gustafsson are with the Department of Electrical Engineering, Link\"{o}ping University, 58183 Link\"{o}ping, Sweden, (e-mail:
tohid@isy.liu.se; emre@isy.liu.se, fredrik@isy.liu.se).
This work is supported by Swedish research council (VR), project scalable Kalman filters.
}
\thanks{ U. Orguner is  with Department of Electrical and Electronics Engineering, Middle East Technical University, 06531 Ankara, Turkey, (email: umut@metu.edu.tr) .}
}
\begin{document}
\maketitle
\begin{abstract}
We present an adaptive smoother for linear state-space models with unknown process and measurement noise covariances. The proposed method utilizes the variational Bayes technique to perform approximate inference. The resulting smoother is computationally efficient, easy to implement, and can be applied to high dimensional linear systems. The performance of the algorithm is illustrated on a target tracking example.
\end{abstract}

\section{Introduction}
Model uncertainties directly affect the performance in filtering and smoothing problems which demand an accurate knowledge of true model parameters. In most practical cases, one's knowledge about the model may not represent the true system. Kalman filters \cite{Kalman60}, which have been widely used in many applications, also require full knowledge of model parameters for reliable estimation. The same requirement is inherited in smoothing methods which use Kalman filters as their building blocks such as Rauch-Tung-Striebel (RTS) smoother \cite{RTS-1965}. 
The sensitivity of the Kalman filter to model parameters has been studied in \cite{Heffes1966,Sangsuk1990,anderson2012} and extensive research is dedicated to the identification of the parameters\cite{Mehra1970,Mehra72,shumway1982EM}.
Noise statistics parameters are of particular interest since they determine the reliability of the information assumed to be hidden in the measurements and the system dynamics.

In the Bayesian approach, one can define priors on the unknown noise parameters and try to compute their posteriors. Here, we use variational approximation for computation of such posteriors where an analytical solution does not exist.
Variational inference based techniques have been used for filtering and smoothing in a number of recent studies.
For example, \cite{Frigola2014} has proposed a procedure for variational Bayesian learning of nonparametric nonlinear state-space models based on sparse Gaussian processes. In the proposed procedure the noise covariances are treated as hyper-parameters and are found via a gradient descent optimization.
 Variational Bayesian (VB) expectation maximization is used in \cite{Beal03} to identify the parameters of linear state-space models, where the process noise covariance matrix is set to the identity  matrix and the remaining parameters are identified up to an unknown transformation.
In \cite{Sarkka09}, the measurement noise covariance is modeled as a diagonal matrix whose entries are assumed to be distributed as inverse Gamma. This result is extended and used in interactive multiple model framework for jump Markov linear systems in \cite{Li11}. In \cite{agamennoni12}, the conjugacy of the inverse Wishart distributed  prior with Gaussian likelihood is exploited to model and estimate the measurement noise covariance in the VB framework. It is also shown in \cite{agamennoni12} that the mean square error of state estimates can be reduced by using the proposed VB measurement update for robust filtering and smoothing. In \cite{Piche12}, the robust filtering and smoothing for nonlinear state-space models with t-distributed measurement noise are given.
In \cite{Barber2007} the parameters of a state-space model and the noise parameters are identified using VB. Although, identification of non-diagonal noise covariances using inverse Wishart distributions is mentioned in \cite{Barber2007}, neither the analysis nor the expressions for the approximate posterior of the inverse Wishart distributed noise covariances are given.
The smoothing under parameter uncertainty can also be cast into an optimization problem; Examples of recent algorithms for robust smoothing for nonlinear state-space models are presented in \cite{agamennoni2014,aravkinopt2013,aravkint2013,aravkinlaplace, Simo2013}. Such optimization based approaches can be used to compute both maximum a posteriori (MAP) and maximum likelihood (ML) estimates of the states and parameters. When the ML estimate is desired, Expectation-Maximization (EM) \cite{EM1977} method can be used as in \cite{shumway1982EM,Ninness2005,Ghahramani96,Simo2013} to compute the ML point  estimate of the noise covariance matrices. In comparison to EM, the VB method, approximates the posterior distribution of the unknown noise parameters and the state variables instead of providing only a point estimate. Further information can be extracted from the posterior as well as the point estimates with respect to different criterion.
In econometrics literature concerning multivariate stochastic volatility such as \cite{Fan2008}, the estimation of covariance matrices is discussed. 

In this letter, we present a novel smoothing algorithm for joint estimation of the state, measurement noise and process noise covariances using the VB technique~\cite[Ch. 10]{Bishop2007},\cite{TzikasLG2008}. Such estimation problems arise when the parameters of a state-space model are found via physical modeling of a system but the noise covariances are unknown.   Our contribution is closely related to \cite{Barber2007}. However, we consider a more general case where both of the noise covariance matrices can be non-diagonal and time-varying.


\vspace{-1mm}
\section{Problem Definition}
\label{sec:pd}
Consider the following linear time-varying state-space representation,
\begin{subequations}
\begin{align}
x_{k+1}=&\ A_kx_k+w_k, & w_k&\stackrel{iid}{\sim}\N(w_k;0,Q_k),\\
y_{k}=&\ C_kx_k+v_k, & v_k&\stackrel{iid}{\sim}\N(v_k;0,R_k), \label{eq:measurementmodel}
\end{align}
\end{subequations}
where $\{x_k\in\mathbb{R}^{n_x}| 0\leq k \leq K\}$ is the state trajectory, also denoted as $x_{0:K}$; $\{y_k\in\mathbb{R}^{n_y}| 0\leq k \leq K\}$  is the measurement sequence,  denoted in more compact form as $y_{0:K}$; $A_k\in\mathbb{R}^{n_x\times n_x}$ and $C_k\in\mathbb{R}^{n_y\times n_x}$ are known state transition and measurement matrices, respectively; $\{w_k\in\mathbb{R}^{n_x}| 0\leq k \leq K-1\}$  and $\{v_k\in\mathbb{R}^{n_y}| 0\leq k \leq K\}$  are mutually independent and white Gaussian noise sequences. The initial state $x_0$ is assumed to have a Gaussian prior, i.e., $p(x_0)=\N(x_0;m_{0},P_{0})$, where
$\N(\cdot;\mu,\Sigma)$ denotes the Gaussian probability density function  (PDF) with mean $\mu$ and covariance $\Sigma$.
$Q_k$ and $R_k$ are the unknown positive definite process noise and measurement noise covariance matrices assumed to have initial inverse Wishart priors
\begin{subequations}
\label{eq:RQ-priors}
\begin{align}
p(Q_0)=&\ \IW(Q_0;\nu_0,V_0),\\
p(R_0)=&\ \IW(R_0;\mu_0,M_0).
\end{align}
\end{subequations}
The inverse Wishart PDF  we use in this work is given in the following form
\begin{equation}
\IW(\Sigma;\nu,\Psi)\triangleq \frac{|\Psi|^{\frac{1}{2}(\nu-d-1)}\exp\tr\left(-\frac{1}{2}\Psi\Sigma^{-1}\right)}{2^{\frac{1}{2}(\nu-d-1)d}\Gamma_d\left[\frac{1}{2}(\nu-d-1)\right]|\Sigma|^{\frac{\nu}{2}}},
\end{equation}
where $\Sigma$ is a  symmetric positive definite random matrix of dimension $d\times d$, $\nu>2d$ is the scalar degrees of freedom and $\Psi$ is a symmetric positive definite matrix of dimension $d\times d$ and is called the scale matrix. This form of the inverse Wishart distribution is  used in~\cite{GuptaN:2000}.
When $\Sigma\thicksim\IW(\Sigma;\nu,\Psi)$, then $\Sigma^{-1}\thicksim\W(\Sigma;\nu-d-1,\Psi^{-1})$ and $\E{[\Sigma]}=\frac{\Psi}{\nu-2d-2}$ when $\nu-2d-2>0$ and $\E[\Sigma^{-1}]={\Psi^{-1}}{(\nu-d-1)}$. Further, any diagonal element of an inverse Wishart matrix is distributed as inverse gamma~\cite[Corollary 3.4.2.1]{GuptaN:2000}. Therefore, the proposed inverse Wishart model is more general than models with diagonal covariance assumption where the diagonal entries are inverse gamma distributed, see e.g.~\cite{Sarkka09}.

It is common in Kalman filtering and smoothing literature  to assume that the noise covariances are fixed parameters \cite{Kailath2000}. However, the noise covariances can be unknown and time-varying. In such cases, the noise parameters can be treated as state variables with dynamics.  
  Dynamical models for covariance matrices   is adopted  here from \cite{caravalho2007} where the matrix Beta-Bartlett stochastic evolution model  was proposed for estimating the multivariate stochastic volatility. The dynamical models for the covariance matrices $R_k$ and $Q_k$  are parametrized by the covariance discount factors $0\ll\lambda_R\leq 1$  and $0\ll\lambda_Q\leq 1$, respectively. 
	The matrix Beta-Bartlett stochastic evolution model  for a generic random  matrix $\Sigma_k$ having a  covariance discount factor $0\ll\lambda\leq 1$ is described in the following.
 Let  $p(\Sigma_{k-1})=\IW(\Sigma_{k-1};\nu_{k-1},\Psi_{k-1})$. The forward predictive  model $p(\Sigma_{k}|\Sigma_{k-1})$ is such that, the forward prediction marginal density becomes the inverse Wishart density parametrized by $p(\Sigma_{k})=\IW(\Sigma_{k};\nu_{k},\Psi_{k})$ where
\begin{subequations}
\begin{align}
\Psi_k&= \lambda \Psi_{k-1},\\
\nu_k&=\lambda \nu_{k-1}+(1-\lambda)(2d+2).
\end{align}
\end{subequations}
Similar to the Kalman filter's prediction update, the forward prediction keeps the marginal expected value of $\Sigma_k$ unchanged while the spread is increased.
Furthermore, the backwards smoothing recursion is given by \cite{caravalho2007}
\begin{subequations}
\begin{align}
\Psi_k^{-1}&\gets (1-\lambda) \Psi_k^{-1}+ \lambda \Psi_{k+1}^{-1},\\
\nu_k&\gets(1-\lambda) \nu_{k}+\lambda\nu_{k+1}.
\end{align}
\end{subequations}
Note that in the prediction and smoothing iterations above, setting $\lambda=1$ corresponds to the fixed parameter case. The aforementioned dynamical model for noise covariances is adopted in  \cite{SimoICML} for adaptive Kalman filtering framework and in \cite{agamennoni12} for filtering and smoothing\footnote{The expressions given in  lines 21 and 22 of Algorithm 3 in \cite{agamennoni12} are inaccurate.
The correct version is given in \cite{caravalho2007}.
} with heavy-tailed measurement noise covariance.

Our aim is to obtain an analytical approximation of the posterior density for the state trajectory $x_{0:K}$ and noise covariances $R_{0:K}$ and $Q_{0:K-1}$. We will derive an approximate smoother which will propagate the sufficient statistics of the approximate distributions through fixed point iterations with guaranteed convergence.

\section{Variational Solution}
\label{sec:mu}
The \textit{a priori} for the joint density $p(x_{0},Q_0,R_0)$ is given as follows,
\begin{align}
p(x_{0},Q_0,R_0)=& \N(x_{0};m_{0},P_{0})\IW(Q_0;\nu_0,V_0)\nonumber\\
&\times \IW(R_0;\mu_0,M_0).
\end{align}
Then,  the posterior  for the state trajectory and the unknown parameters denoted by $\Z \triangleq \{x_{0:K},Q_{0:K-1},R_{0:K}\}$ is given by Bayes' theorem as
\begin{align}
p(\Z|&y_{0:K})\propto p(x_{0},Q_0,R_0)p(y_K|x_K,R_K)\prod_{l=0}^{K-2}  p(Q_{l+1}|Q_l)  \nonumber\\
&\times \prod_{k=0}^{K-1} p(y_k|x_k,R_k)  p(x_{k+1}|x_k,Q_k) p(R_{k+1}|R_k).
\end{align}
There is no analytical solution for this posterior. We are going to look for an approximate analytical solution using the following variational approximation.
\begin{align}
p(\Z|y_{0:K})&\approx q(\Z)\triangleq q_x(x_{0:K}) q_{Q}(Q_{0:K-1}) q_{R}(R_{0:K}),\label{eq:factors}
\end{align}
where the densities $q_x(\cdot)$, $q_{Q}(\cdot)$ and $q_{R}(\cdot)$ are the approximate posterior densities for $x_{0:K}$, $Q_{0:K-1}$ and $R_{0:K}$, respectively.
The well-known technique of VB~\cite[Ch. 10]{Bishop2007},\cite{TzikasLG2008} chooses the estimates $\hat{q}_x(\cdot)$, $\hat{q}_{Q}(\cdot)$ and $\hat{q}_{R}(\cdot)$ for the factors in \eqref{eq:factors} using the following optimization problem
\begin{align}
&\hat{q}_{x},\hat{q}_{Q},\hat{q}_{R}=\argmin_{{q}_{x},{q}_{Q},{q}_{R}}{D_{KL}(q(\Z)||p(\Z|y_{0:K}))} \label{eq:KLcost}
\end{align}
where $D_{KL}(q(x)||p(x))\triangleq\int q(x)\log\frac{q(x)}{p(x)}\d x$ is the Kullback-Leibler divergence~\cite{CoverT2006}.
The optimal solution satisfies the following set of equations.
\begin{subequations}
\label{eqn:IterativeOptimization}
\begin{align}
&\log \hat{q}_{x}(x_{0:K})=\E_{\hat{q}_{Q}\hat{q}_{R}}[\log p(\Z,y_{0:K})]+c_{x},\label{eqn:IterativeOptimizationx}\\
&\log \hat{q}_{Q}(Q_{0:K-1})=\E_{\hat{q}_{x}\hat{q}_{R}}[\log p(\Z,y_{0:K})]+c_{Q},\label{eqn:IterativeOptimizationQ}\\
&\log \hat{q}_{R}(R_{0:K})=\E_{\hat{q}_{x}\hat{q}_{Q}}[\log p(\Z,y_{0:K})]+c_{R},\label{eqn:IterativeOptimizationR}
\end{align}
\end{subequations}
where $c_{x}$, $c_Q$ and $c_R$  are constants with respect to the variables $x_{0:K}$, $Q_{0:K-1}$ and $R_{0:K}$,  respectively.
The solution to~\eqref{eqn:IterativeOptimization} can be obtained via fixed-point iterations where only one factor in~\eqref{eq:factors} is updated and all the other factors are fixed to their last estimated values~\cite[Ch. 10]{Bishop2007}. The iterations converge to a local optima of \eqref{eq:KLcost}~\cite[Ch. 10]{Bishop2007}, \cite[Ch. 3]{jordan2008}. The complete (standard but tedious) derivations for the  variational iterations are given in \cite{ArdeshiriRQderivation}.
The implementation pseudo-code for the proposed algorithm is given in Table~\ref{table:smoothing}. When the recursions of the proposed algorithm converge, the expected values or the modes of the posteriors for $x_k$, $R_k$ and $Q_k$ can be used as the point estimates for the random variables. When an estimate of uncertainty for the point estimate is required the posterior variances can be used. Nevertheless, it is well-known that the VB method underestimates the covariance when the posterior is multi-modal~\cite[Ch. 10]{Bishop2007}.
\section{Simulations}
\label{sec:sim}
\subsection{Unknown time-varying noise covariances} \label{sec:numsimtVar}
We illustrate the performance of the proposed smoother in an object tracking scenario.
In the simulation scenario, a point object moves according to the continuous white noise acceleration model in two dimensional Cartesian coordinates~\cite[p.~269]{bar2004estimation}.
The sampling time is $\tau=1s$ and the simulation length $K$ is chosen to be 4000. The state vector is defined as the position and the velocity of the object. A sensor collects noisy measurements of the object's position according to \eqref{eq:measurementmodel}. 
The true parameters of the linear state-space model are given as
\begin{align*}
				 &A_k=\diag(a,a),  & & 	Q_0=\diag(q,q),\\
				& a= \begin{bmatrix} 1 & \tau\\ 0 & 1 \end{bmatrix}, && q= \sigma_\nu^2\begin{bmatrix} \tau^3/3& \tau^2/2\\ \tau^2/2&\tau\end{bmatrix}, \\
				& R_k^{\text{True}}=\left(2-\cos\left(\frac{4\pi k}{K}\right)\right)R_0,  && R_0=\sigma_e^2\begin{bmatrix} 5 &  1\\1&5 \end{bmatrix},\\
				&Q_{k}^{\text{True}}=\left(\frac{2}{3}+\frac{1}{3}\cos\left(\frac{4\pi k}{K}\right)\right) Q_0,&&	C_k=\begin{bmatrix} 1 &  0&0&0\\0&0&1&0 \end{bmatrix}.
\end{align*}
The  noise related parameters  are $\sigma_e^2=2m^2$ and $\sigma_v^2=3m^2/s^{3}$.
The initial values of the parameters at time index $k=0$ are used in the RTS smoother which are $R_{0}$ and $Q_{0}$, respectively.
Using the simulated measurement data, we compare four smoothers; RTS smoother using the fixed noise covariances $R_0$ and $Q_0$ (denoted by RTS), VB smoother for estimating only $R_k$ as given in \cite[Algorithm 3]{agamennoni12} (denoted by VBS-R), the proposed VB algorithm for estimating $R_k$ and $Q_k$ simultaneously  (denoted by VBS-RQ),
and the oracle RTS smoother which knows the true noise covariances (denoted by Oracle-RTS).
\begin{table}[t]
\caption{Smoothing with unknown noise covariances}\label{table:smoothing}
\vspace{-5mm}\rule{\columnwidth}{1pt}
\begin{algorithmic}[1]
\State \textbf{Inputs:} $A_k$, $C_k$, $\nu_0$, $V_0$, $\mu_0$, $M_0$, $m_0$,  $P_0$, $\lambda_Q$, $\lambda_R$ and $y_{0:K}$.
\Statex \hspace{0mm}\textit{initialization}
\State $V_{k|K}\gets V_0$, $\nu_{k|K}\gets \nu_0$, $M_{k|K}\gets M_0$, $\mu_{k|K}\gets \mu_0$ for $0\leq k\leq K$
\Repeat
\Statex \hspace{2mm}\textit{update $q_x(x_{0:K})$ given $q_Q(Q_{0:K-1})$ and $q_R(R_{0:K})$}
	\State $\widetilde{Q}_{k}\gets V_{k|K}/(\nu_{k|K}-n_x-1)$ for $0\leq k\leq K$
	\State $\widetilde{R}_k\gets M_{k|K}/(\mu_{k|K}-n_y-1)$ for $0\leq k\leq K$
	\State $m_{0|-1}\gets m_0$, $P_{0|-1}\gets P_0$
	\For{$k$ = 0 to K }
		\State $K_k\gets P_{k|k-1}C_k^\t(C_kP_{k|k-1}C_k^\t+\widetilde{R}_k)^{-1}$
		\State $m_{k|k}\gets m_{k|k-1}+K_k(y_k-C_km_{k|k-1})  $
		\State $P_{k|k}\gets (I-K_kC_k)P_{k|k-1}$
	  \State $m_{k+1|k} \gets A_km_{k|k}$
	  \State $P_{k+1|k} \gets A_kP_{k|k}A_k^\t+\widetilde{Q}_k$	
	\EndFor
	\For{$k$ = K-1 down to 0 }
		\State $G_k\gets P_{k|k}A_k^\t P_{k+1|k}^{-1}$
		\State $m_{k|K}\gets m_{k|k}+G_k(m_{k+1|K}-A_km_{k|k})$
		\State $P_{k|K}\gets P_{k|k}+G_k(P_{k+1|K}-P_{k+1|k})G_k^\t$
	\EndFor
	\Statex \hspace{2mm}\textit{update $q_Q(Q_{0:K-1})$  and $q_R(R_{0:K})$ given  $q_x(x_{0:K})$}
  \State $\nu_{0|-1}\gets \nu_0$, $V_{0|-1}\gets V_{0}$, $\mu_{0|-1}\gets \mu_0$, $M_{0|-1}\gets M_{0}$		
	\For{$k$ = 0 to K }
			\State $\mu_{k|k}\gets \mu_{k|k-1}+1$
			\State $M_{k|k}\gets M_{k|k-1} +C_kP_{k|K}C_k^\t{}$
			\Statex $ \hspace{15mm} +(y_k-C_km_{k|K})(y_k-C_km_{k|K})^\t$
			\State $\mu_{k+1|k}\gets \mu_{k|k}+(1-\lambda_R)(2n_y+2)$, $M_{k+1|k}\gets \lambda_R M_{k|k}$
	\EndFor
	\For{$k$ = 0 to K-1 }
			\State $\nu_{k|k}\gets \nu_{k|k-1}+1$
			\State $V_{k|k}\gets V_{k|k-1} + P_{k+1|K}+A_{k}P_{k|K}A_{k}^\t{}-P_{k+1,k|K}A_{k}^\t{}$
			\Statex $\hspace{2mm}-A_{k}P_{k,k+1|K}+(m_{k+1|K}-A_{k}m_{k|K})(m_{k+1|K}-A_{k}m_{k|K})^\t{}$
			\State $\nu_{k+1|k}\gets \nu_{k|k}+(1-\lambda_Q)(2n_x+2)$, $V_{k+1|k}\gets \lambda_Q V_{k|k}$
	\EndFor
	\For{$k$ = K-1 down to 0 }
			\State  $\mu_{k|K}\gets(1-\lambda_R) \mu_{k|k} + \lambda_R \mu_{k+1|K}$
			\State $M_{k|K}\gets \left((1-\lambda_R) M_{k|k}^{-1}+ \lambda_R M_{k+1|K}^{-1}\right)^{-1}$
	\EndFor
	\For{$k$ = K-2 down to 0 }
			\State  $\nu_{k|K}\gets(1-\lambda_Q) \nu_{k|k} + \lambda_Q \nu_{k+1|K}$
			\State  $V_{k|K}\gets \left((1-\lambda_Q) V_{k|k}^{-1}+ \lambda_Q V_{k+1|K}^{-1}\right)^{-1}$
	\EndFor
\Until{\textbf{converged}}
\State \textbf{Outputs:} $m_{k|K}$,  $P_{k|K}$, $M_{k|K}$, $\mu_{k|K}$ and,
\Statex  $\widehat{R_k} \triangleq\E_{q_R}[R_k|y_{0:K}]={M_{k|K}}/({\mu_{k|K}-2n_y-2})$ for $k=0\cdots K$.
\Statex  $V_{k|K}$, $\nu_{k|K}$ and $\widehat{Q_k} \triangleq\E_{q_Q}[Q_k|y_{0:K}]={V_{k|K}}/({\nu_{k|K}-2n_x-2})$  for  $k=0\cdots K-1$.
\end{algorithmic}
\noindent \rule{\columnwidth}{1pt}\vspace{-6mm}
\end{table}

We perform  $N_{MC}=5000$ Monte Carlo (MC) simulations. In each MC run, a new trajectory with an initial state and the corresponding measurements are generated.
The prior for the initial state is assumed to be Gaussian, i.e.,  $x^j_0 \stackrel{iid}{\sim} p(x_0)=\N(x_0;m_{0},P_{0})$ for the $j^{\text{th}}$ MC simulation where,
\begin{subequations}
\begin{align}
m_{0}&=[0m,5m/s,0m,5m/s]^\t{},\\
P_{0}&=\diag([30^2, 30^2, 30^2,30^2]).
\end{align}
\end{subequations}
The initial parameters of the inverse Wishart prior densities in \eqref{eq:RQ-priors} for the smoother VBS-RQ  are chosen as
\begin{subequations}
\begin{align}
\nu_0&= 2n_x+3, & V_0&=(\nu_0-2n_x-2)Q_{0}, \\
\mu_0&= 2n_y+3, & M_0&=(\mu_0-2n_y-2)R_{0}. \label{eq:muM}
\end{align}
\end{subequations}
This choice of initial parameters yields the expected value of the initial prior densities of $R_k$ and $Q_k$ to coincide with the nominal values of $R_0$ and $Q_0$. Similarly, the initial parameters of the inverse Wishart prior density for VBS-R are given in \eqref{eq:muM}. In the VBS-RQ and VBS-R, covariance discount factors are set to $\lambda=0.98$ and, the number of iterations in the variational update is set to $50$.

We compare the four smoothers in terms of the root mean square error (RMSE) of the position estimates
\begin{align}
\text{RMSE}(j) &\triangleq \left ( \frac{1}{K+1} \sum_{k=0}^{K} {\left\| C(m_{k|K}^{j}-x_k^{j}) \right\|_2^2} \right)^{\frac{1}{2}} \label{eq:RMSE}.
\end{align}
and its average RMSE over all MC simulations denoted by ARMSE. In \eqref{eq:RMSE}, $m_{k|K}^{j}$ and $x_k^{j}$ denote the estimated mean of state $x_k$ and its true value in the $j^{\text{th}}$ MC run, respectively.
For the matrices, we use the square root of the average Frobenius norm square normalized by the number of elements as the error measure
\begin{subequations}
\label{eq:ERQ}
\begin{align}
E_{R}(j) &\triangleq \left( \frac{1}{n_y^2 (K+1)} \sum_{k=0}^{K}{\tr \left((\widehat{R_{k}}^{j}-R_k{^\text{True}})^2\right) }\right)^{\frac{1}{4}} \label{eq:E_R}\\
E_{Q}(j) &\triangleq \left( \frac{1}{n_x^2 K} \sum_{k=0}^{K-1}  {\tr \left((\widehat{Q_k}^{j}-Q_k{^\text{True}})^2\right) }\right)^{\frac{1}{4}} \label{eq:E_Q}.
\end{align}
\end{subequations}
In \eqref{eq:ERQ}, $\widehat{R_{k}}^{j}$, $\widehat{Q_k}^{j}$, $R_k{^\text{True}}$ and $Q_k{^\text{True}}$ denote the estimated mean of measurement noise covariance, process noise covariance and their true values in the $j^{\text{th}}$ MC run, respectively.

\begin{figure}[t]
  \centering
  \subfigure[]{\includegraphics[width=0.45\textwidth]{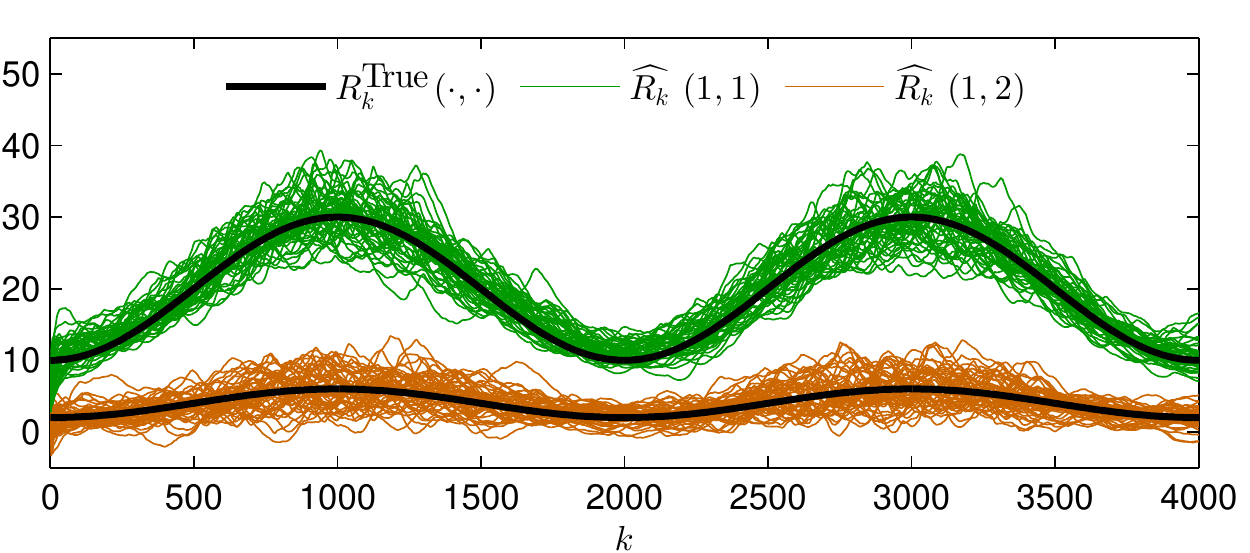}
  \label{fig:RtVar}}
	\subfigure[]{\includegraphics[width=0.45\textwidth]{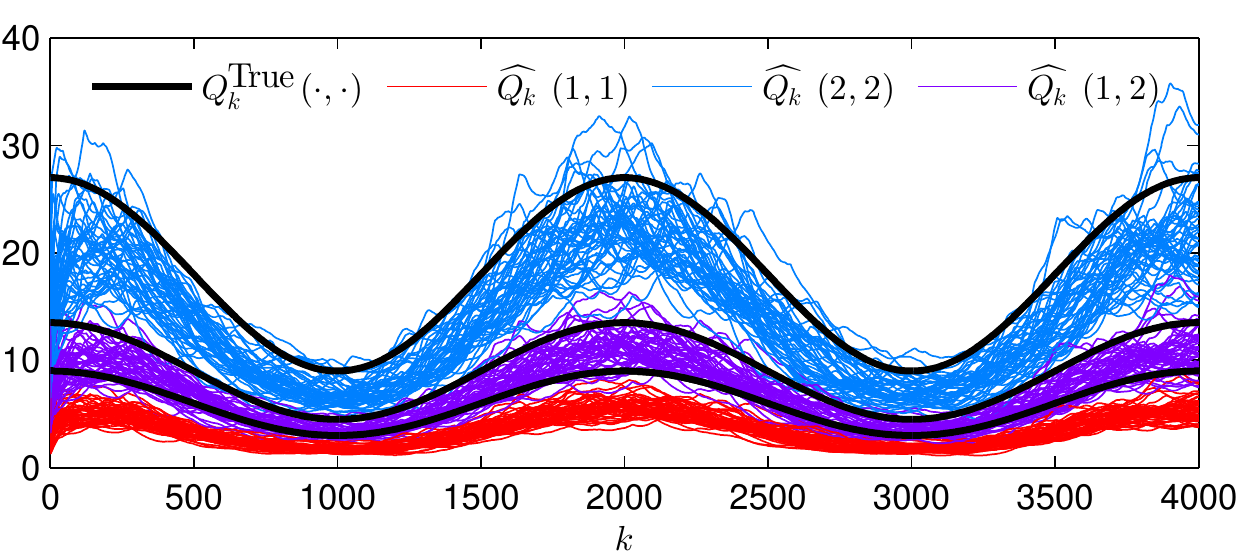}
  \label{fig:QtVar}}
	\caption{ 
	(a) First diagonal  and the off-diagonal elements of the estimated measurement noise covariance $\widehat{R_k}$ using VBS-RQ are plotted versus time index $k$ along with the their true value (in black). (b) First two diagonal  and an off-diagonal element of the estimated process noise covariance $\widehat{Q_k}$ using VBS-RQ are plotted versus time index $k$ along with the their true value (in black).}
	  \label{fig:RQtVar}
		\vspace{-3mm}
\end{figure}

The estimates of some elements of the  noise covariances versus time for some random samples of MC runs along with their true value are given in Fig.~\ref{fig:RQtVar}. 

Error corresponding to the value for $R_0$ and $Q_0$ used in RTS computed using \eqref{eq:ERQ} are $E_R=2.972$ and $E_Q=2.224$, respectively. The error values for the smoothers  are  given in Table~\ref{table:VBRQtVar}.
For those algorithms which do not estimate $R_k$ and $Q_k$ the corresponding error terms are not given in Table~\ref{table:VBRQtVar}.
\begin{table}[t]
\caption{Time-varying noise covariances:  Comparison of four smoothers in terms of estimation errors.}
\centering
\scalebox{0.8}{
\begin{tabular}{l ccc }
\toprule
  &\multicolumn{3}{c} {\textbf{Errors (Mean $\pm$ Standard deviation)}}    \\ 
\cmidrule(lr){2-4} 
\textbf{Smoothers} &\textbf{RMSE}&\textbf{$E_R$}&\textbf{$E_Q$}\\\midrule
\textbf{Oracle-RTS}&3.608$\ \pm\ $0.045&--&--\\
\textbf{RTS}&3.879$\ \pm\ $0.047&--&--\\
\textbf{VBS-R}&3.712$\ \pm\ $0.047&1.687$\ \pm\ $0.079&--\\
\textbf{VBS-RQ}&3.653$\ \pm\ $0.047&1.485$\ \pm\ $0.070&1.572$\ \pm\ $0.063\\
\bottomrule
\end{tabular}
\label{table:VBRQtVar}

}
\vspace{-3mm}
\end{table}
\subsection{Unknown time-invariant noise covariances}
When the noise covariances are time-invariant unknown parameters, the EM algorithm \cite[page 182]{Simo2013} offers an  alternative to the proposed VB algorithm. An EM algorithm for smoothing with unknown noise covariances (denoted by EMS-RQ) is given in \cite{ArdeshiriRQderivation} and is compared to VBS-RQ.
 Here, we will repeat the simulation scenario in Section~\ref{sec:numsimtVar} for $K=1000$ and compare six smoothers; RTS, VBS-R,  VBS-RQ, Oracle-RTS, EMS-RQ and the VB smoother with diagonal covariance matrices with inverse gamma distributed entries proposed in \cite{Barber2007} (denoted by VBS-RQ-D).  Since the noise covariances are now fixed parameters, we set $\lambda=1$ in VBS-RQ. Furthermore, we drop the time index $k$ and denote the noise covariances by $R$ and $Q$ in the rest of this section.

\begin{figure}[t]
  \centering
\subfigure[]{\includegraphics[width=0.45\textwidth]{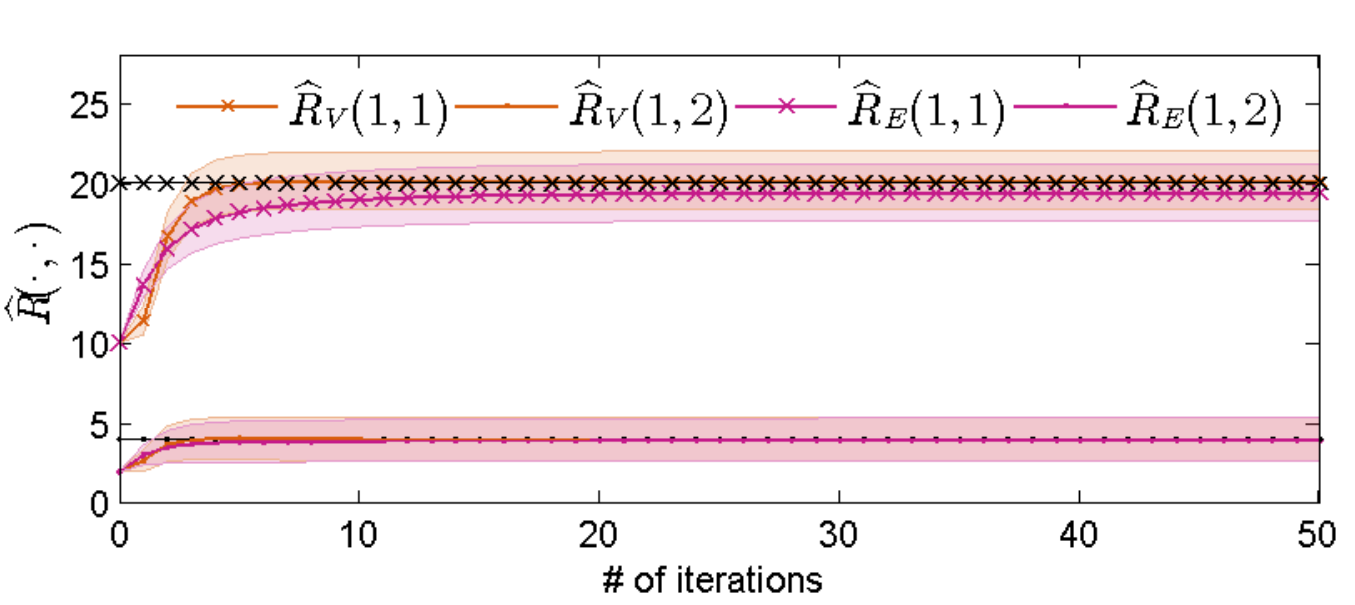}
\label{fig:VBRQdiagonalsR}}
\subfigure[]{\includegraphics[width=0.45\textwidth]{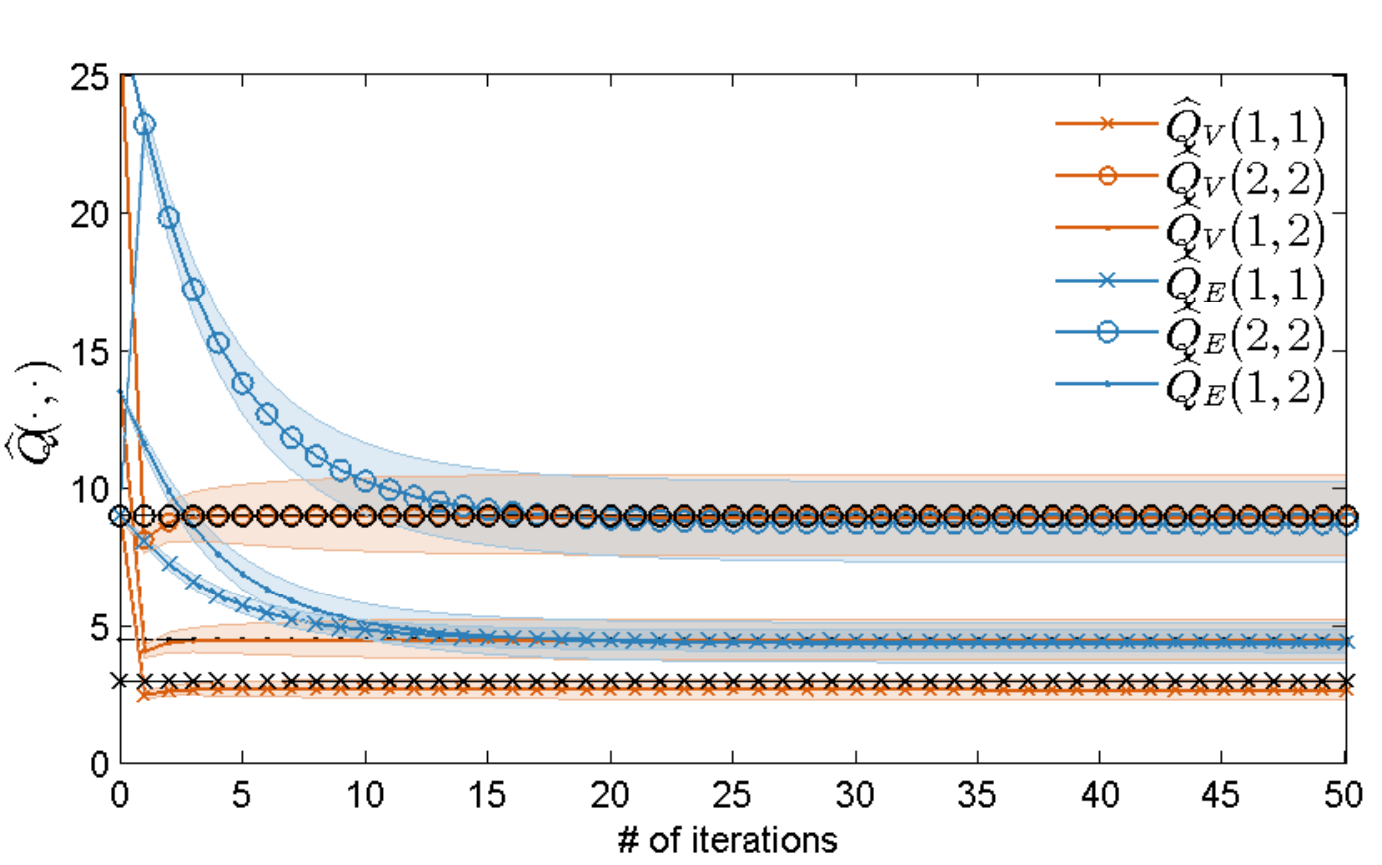}
\label{fig:VBRQdiagonalsQ}}
\caption{ Estimated elements of noise covariance matrices using VBS-RQ (denoted by subscript $V$) and EMS-RQ (denoted by subscript $E$) are plotted versus number of iterations of the algorithm along with their true corresponding value (in black).  The median values are plotted is solid line along with  shaded areas in the same color as the median curve illustrating the interval between  $5$ and $95$ percentiles.
(a) First  diagonal element and the off-diagonal element of estimated measurement noise covariance matrices.
(b) First two diagonal elements and an off-diagonal element of estimated process noise covariance matrices.}
\label{fig:VBRQdiagonals}
\end{figure}

The nominal values of the parameters used in the RTS smoother are $R_{0}$ and $Q_{0}$, respectively while their true values are  $R^{\text{True}}=2 R_0$ and $Q^{\text{True}}=0.2 Q_0$.
Error corresponding to the nominal value for $R$ and $Q$ computed using \eqref{eq:ERQ} are $E_R=2.685$ and $E_Q=2.842$, respectively. The error values for the smoothers  are  given in Table~\ref{table:VBRQ}.
For those algorithms which do not estimate $R$ and $Q$ the corresponding error terms are not given in Table~\ref{table:VBRQ}.  
The convergence of some elements of the $\widehat{R}$ and $\widehat{Q}$ versus the number of iterations is illustrated in Fig.~\ref{fig:VBRQdiagonals}.
\begin{table}[t]
\caption{Time-invariant  noise covariances:Comparison of six smoothers in terms of estimation errors.}
\centering
\scalebox{0.8}{
\begin{tabular}{l ccc }
\toprule
  &\multicolumn{3}{c} {\textbf{Errors (Mean $\pm$ Standard deviation)}}    \\ 
\cmidrule(lr){2-4} 
\textbf{Smoothers} &\textbf{RMSE}&\textbf{$E_R$}&\textbf{$E_Q$}\\\midrule
\textbf{Oracle-RTS}&3.399$\ \pm\ $0.088&--&--\\
\textbf{RTS}&3.786$\ \pm\ $0.090&--&--\\
\textbf{VBS-R}&3.595$\ \pm\ $0.090&1.326$\ \pm\ $0.195&--\\
\textbf{VBS-RQ}&3.402$\ \pm\ $0.088&0.929$\ \pm\ $0.211&0.668$\ \pm\ $0.129\\
\textbf{EMS-RQ}&3.407$\ \pm\ $0.088&0.975$\ \pm\ $0.212&0.851$\ \pm\ $0.075\\
\textbf{VBS-RQ-D}&3.433$\ \pm\ $0.089&1.764$\ \pm\ $0.056&1.515$\ \pm\ $0.009\\
\bottomrule
\end{tabular}
\label{table:VBRQ}

}\vspace{-3mm}
\end{table}
\section{Discussion and Conclusion}
\label{sec:con}

We have proposed a smoothing technique based on a variational Bayes approximation. We have shown a successful numerical simulation using variational Bayes for approximate inference for a linear state-space model  with unknown time-varying measurement noise and process noise covariances.
 In our simulations we obtain lower ARMSE  for the state estimate compared to RTS smoother in presence of modeling mismatch. Furthermore,  we obtain lower ARMSE  for the state estimate  compared to  other state-of-the-art smoothers which  identify the noise covariances using EM  which is a consequence of the fact that the algorithm iteratively finds a better estimate of the process noise and measurement noise covariances. 

The proposed algorithm for general time-varying noise covariance estimation can be restricted to a fixed noise parameter estimation algorithm by choosing a unity covariance discount factor. Furthermore, when the sparsity pattern in a covariance matrix is known a priori, the elements which are zero  can be set to zero to obtain a tailored algorithm.

\bibliographystyle{IEEETran}
\bibliography{IEEEabrv,VariationalNoiseAdaptive}

\end{document}